\documentclass{article}
\usepackage{spconf,amsmath,graphicx}
\usepackage{amsfonts}
\usepackage{algorithmic}
\usepackage{algorithm}
\usepackage{array}
\usepackage[caption=false,font=normalsize,labelfont=sf,textfont=sf]{subfig}
\usepackage{textcomp}
\usepackage{stfloats}
\usepackage{url}
\usepackage{verbatim}
\usepackage{cite}
\usepackage{acronym, color, soul}
\usepackage{hyperref}
\usepackage{booktabs}
\usepackage{multirow, bm}
\usepackage{adjustbox}
\usepackage{xspace}

\makeatletter
\DeclareRobustCommand\onedot{\futurelet\@let@token\@onedot}
\def\@onedot{\ifx\@let@token.\else.\null\fi\xspace}

\def\etal{\emph{et al}\onedot}

\newacro{ai}[AI]{artificial intelligence}
\newacro{gpu}[GPU]{graphics processing unit}
\newacro{cpu}[CPU]{central processing unit}
\newacro{convnet}[ConvNet]{convolutional neural network}
\newacro{dnn}[DNN]{deep neural network}
\newacro{ann}[ANN]{artificial neural network}
\newacro{rnn}[RNN]{recurrent neural network}
\newacro{vae}[VAE]{variational autoencoder}
\newacro{ae}[AE]{autoencoder}
\newacro{resblock}[ResBlock]{residual block}
\newacro{gelu}[GELU]{gaussian error linear unit}

\newacro{vvc}[VVC]{versatile video coding}
\newacro{hevc}[HEVC]{high-efficiency video coding}
\newacro{avc}[AVC]{advanced video coding}

\newacro{psnr}[PSNR]{peak signal-to-noise ratio}
\newacro{ms-ssim}[MS-SSIM]{multi-scale structural similarity index}
\newacro{ssim}[SSIM]{structural similarity index}
\newacro{mae}[MAE]{mean absolut error}
\newacro{vc}[VC]{video codec}
\newacro{nvc}[NVC]{neural video coding}
\newacro{nerv}[NeRV]{neural representations for videos}
\newacro{ffnerv}[FFNeRV]{frame-wise neural representations for videos}
\newacro{inr}[INR]{implicit neural representation}
\newacro{nerf}[NeRF]{neural radiance field}
\newacro{mlp}[MLP]{multi-layer perceptron}
\newacro{qat}[QAT]{quantization-aware training}
\newacro{ptq}[PTQ]{post-training quantization}
\newacro{scrb}[SCRB]{separable conv2d residual block}
\newacro{ub}[UB]{upsampling block}
\newacro{rd}[RD]{rate-distortion}
\newacro{flops}[FLOPs]{floating point operations per second}
\newacro{vsr}[VSR]{video super resolution}
\newacro{dvc}[DVC]{deep video compression}
\newacro{mac}[MAC]{multiplication-accumulation}
\newacro{rd}[RD]{rate-distorsion}
\newacro{gop}[GoP]{groupe of frames}
\newacro{lvc}[LVC]{learned video compression}
\newacro{uvg}[UVG]{ultra video group}
\newacro{pe}[PE]{positional encoding}
\newacro{adain}[AdaIN]{adaptive instance normalization}
\newacro{ssf}[SSF]{scale-space flow}
\newacro{convnet}[ConvNet]{convolution network}

\title{NeRV++: An Enhanced Implicit Neural Video Representation}

\name{Ahmed Ghorbel$^{1}$ \qquad  Wassim Hamidouche$^{1,2}$ \qquad Luce Morin$^{1}$}
\address{%
$^1$ Univ Rennes, INSA Rennes, CNRS, IETR – UMR 6164, F-35000 Rennes, France \\
$^{2}$ Technology Innovation Institute P.O.Box: 9639, Masdar City Abu Dhabi, UAE}

\begin{document}
\maketitle

\begin{abstract}
Neural fields, also known as \acp{inr}, have shown a remarkable capability of representing, generating, and manipulating various data types, allowing for continuous data reconstruction at a low memory footprint.
Though promising, \acp{inr} applied to video compression still need to improve their rate-distortion performance by a large margin, and require a huge number of parameters and long training iterations to capture high-frequency details, limiting their wider applicability. Resolving this problem remains a quite challenging task, which would make \acp{inr} more accessible in compression tasks.
We take a step towards resolving these shortcomings by introducing \ac{nerv}++, an enhanced implicit neural video representation, as more straightforward yet effective enhancement over the original \ac{nerv} decoder architecture, featuring \acp{scrb} that sandwiches the \ac{ub}, and a bilinear interpolation skip layer for improved feature representation. \ac{nerv}++ allows videos to be directly represented as a function approximated by a neural network, and significantly enhance the representation capacity beyond current \ac{inr}-based video codecs.
We evaluate our method on UVG, MCL\_JVC, and Bunny datasets, achieving competitive results for video compression with \acp{inr}. This achievement narrows the gap to autoencoder-based video coding, marking a significant stride in \ac{inr}-based video compression research.
\end{abstract}


\acresetall

\section{Introduction}
\label{intro}
%
%
The proliferation of digital video content in various domains, ranging from entertainment and education to surveillance and remote communication, has spurred intensive research into efficient video processing and compression techniques.
\begin{figure}[htb]
\centering
\includegraphics[width=0.49\textwidth]{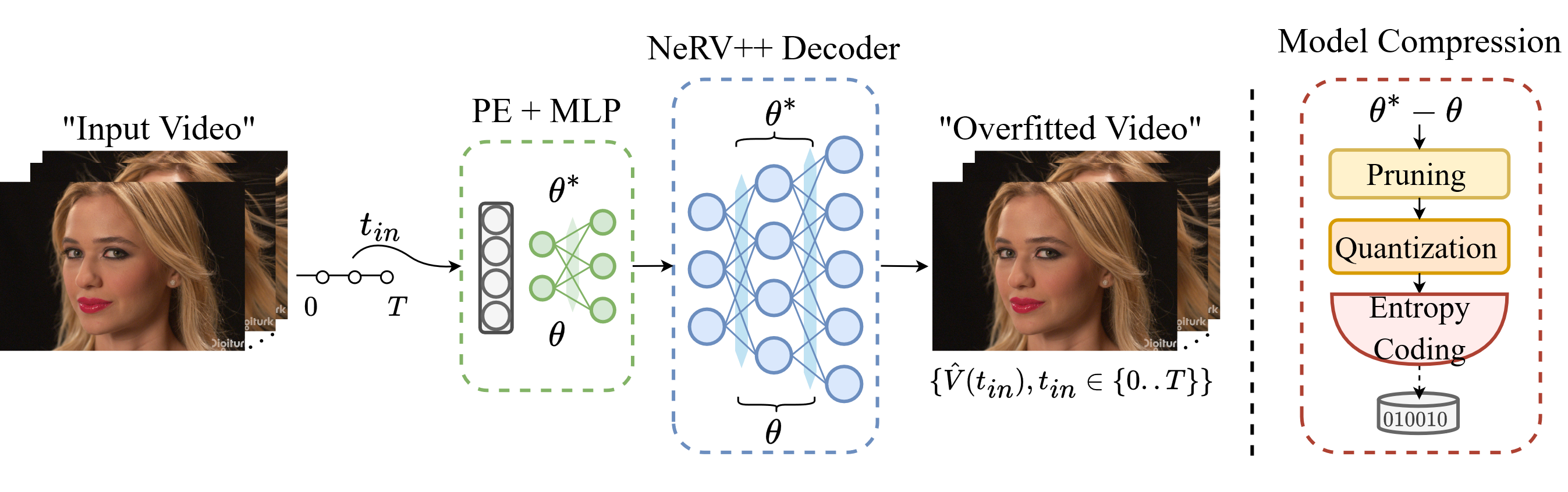}
\caption{High-level diagram of implicit neural representation for video compression.}
\label{HL_diag}
\end{figure}
In this era, conventional methods, based on established video coding standards such as \ac{hevc}/H.265~\cite{6316136} and \ac{vvc}/H.266~\cite{9503377}, have made notable progress in minimizing spatial and temporal redundancies, facilitating efficient storage and transmission of video data. Nevertheless, the growing demand for high-quality video streaming, real-time applications, and immersive experiences requires the exploration of innovative approaches to video representation.

In response to these challenges, the advent of \ac{lvc} techniques, driven by \acp{dnn}, has ushered in a paradigm shift. These methods leverage the power of \ac{ai} to learn intricate patterns within video data, enabling adaptive compression schemes tailored to the content's complexity. Several data-driven attempts utilizing \ac{convnet} and autoencoder architectures to compress video signals have been made~\cite{rippel2019learned,lu2019dvc}. These approaches assume that shared decoder networks only need to store or transmit essential video information. Specifically, the encoder generates latent representations of a sequence of frames, which are then quantized and entropy-coded. The decoder subsequently reconstructs an approximate version of the original sequence. This approach often includes temporal consistency priors by selecting key frames within a group of frames and encoding information to reconstruct the remaining frames separately. Several works build on this approach, proposing different forms of representation for the key and predicted frames~\cite{mentzer2022vct,djelouah2019neural,agustsson2020scale}. For instance, the \ac{dvc} method proposed by Lu \etal~\cite{lu2019dvc} has demonstrated compression ratios comparable to or slightly better than software implementations x264 and x265 of \ac{avc}/H.264~\cite{richardson2011h} and \ac{hevc}/H.265~\cite{sullivan2012overview} standards. This redefines the compression pipeline using neural networks instead of hand-crafted algorithms. However, these autoencoder-based approaches are inherently susceptible to biases present in their training datasets and pose challenges due to their extensive parameterization. Learned decoders, often comprising millions of parameters and requiring up to a million \ac{mac} operations to decode a single pixel, are several orders of magnitude more complex than conventional counterparts. This complexity poses a potential hindrance to their widespread adoption, despite their promising compression capabilities~\cite{rippel2019learned,lu2019dvc}.

In recent years, \acp{inr}, particularly models like DeepSDF~\cite{park2019deepsdf}, \ac{nerf}~\cite{mildenhall2021nerf}, and their derivatives, have played a crucial role in revolutionizing 3D object modeling and image synthesis. They have demonstrated significant success in 3D reconstruction~\cite{sucar2021imap,zhu2022nice} and novel view synthesis applications~\cite{martin2021nerf,meng2021gnerf,muller2022instant,zhang2020nerf++}, leveraging their compactness and expressiveness. Implemented through neural networks, these \acp{inr} offer continuous signal representations, providing versatility by allowing training directly on samples from the signal to be fit (e.g., mapping $(x, y)$ coordinates to $RGB$ color values for images) or through outputs from differentiable processes, as exemplified by \ac{nerf}. The capability to parameterize continuous 3D representations positions \acp{inr} as a pivotal technology in modern computational 3D and image processing fields. However, these networks encounter the spectral bias issue, struggling to capture high-frequency details. Solutions like SIREN, proposed by Sitzmann \etal~\cite{sitzmann2020implicit}, use periodic activation functions to mitigate this issue, offering improved representation of images and videos. Additionally, Mildenhall \etal~\cite{mildenhall2021nerf} applied positional encoding, mapping input coordinates to higher dimensions using sinusoidal functions, thereby enhancing the network's capability to capture fine details. Our work also integrates positional encoding, aligning with recent advancements in \ac{inr}-based video compression.
In the realm of image compression using \ac{inr}, an image is conceptualized as a function $f(x, y) = (R, G, B)$, approximated by fitting a neural network to a set of pixels $P = {(x, y), (R, G, B)}$. The image is then effectively stored in the neural network’s parameters and can be recovered by performing forward passes. Notably, the use of \ac{inr} frames image compression as a neural network compression problem. The pioneering work by Dupont \etal~\cite{dupont2021coin} in applying \ac{inr} to image compression marked a significant advancement in this field. Leveraging SIREN-based models, Sitzmann \etal~\cite{sitzmann2020implicit} demonstrated the efficiency of \ac{inr} over JPEG in low-bitrate scenarios. Their approach involved fitting a neural network to an image and compressing it through post-training quantization. Strümpler \etal~\cite{strumpler2022implicit} and Dupont \etal~\cite{dupont2022coin++} furthered this research by introducing a meta-learned model at the receiver side and explicit quantization methods. These approaches achieved favorable results by learning and transmitting only the network modulations, then fine-tuning for performance recovery. Their research also explored the implementation of $L1$ regularization to potentially reduce entropy within the weights. However, adjusting the intensity of this regularization displayed only negligible changes in the efficiency of the compressed output. This observation led to the conclusion that $L1$ regularization might not effectively serve as a mechanism for minimizing entropy in such contexts.

All the aforementioned \ac{inr} approaches are pixel-based implicit representations, when handling extensive and high-resolution datasets, they exhibit limitations in training and testing time efficiency, rendering them suboptimal for various applications. Conversely, \ac{nerv}~\cite{chen2021nerv} introduces an image-based implicit representation strategy, integrating convolutional operations with \ac{inr} to jointly learn an \ac{inr} across all pixel values. This approach enhances data processing speed and streamlines model training. While this approach sacrifices spatial continuity, it enables the use of convolutional layers, which are more prevalent and effective in image processing tasks compared to fully connected layers. Thus, video encoding in \ac{nerv}~\cite{chen2021nerv} involves fitting a neural network to frames, with the decoding being a straightforward feedforward process. Initially, the network undergoes training to minimize distortion loss, followed by procedures aimed at diminishing its size, such as converting its weights to an 8-bit floating-point format, alongside quantization and pruning techniques. Their results show that image-based neural video representations can provide superior performance than pixel-based representations for the task of video compression. Additionally, training and inference times are significantly reduced.


Li \etal~\cite{li2022nerv} furthered the \ac{nerv}~\cite{chen2021nerv} approach and proposed E-\ac{nerv}, an optimization to this architecture, by segregating spatial and temporal input coordinates. Thereby reducing reliance on large fully connected layers required by \ac{nerv}, resulting in a more efficient allocation of network parameters. Recently, Bai \etal proposed PS-\ac{nerv}~\cite{bai2023ps}, a patch-wise approach to video representation using \acp{inr}, including \ac{adain} for feature enhancement and capturing high-frequency details. Further, lee \etal \cite{lee2022ffnerv} proposed FF\ac{nerv} that leverages optical flows to exploit temporal redundancies, and adopted multi-resolution temporal grids for mapping continuous temporal coordinates and employing compact convolutional designs with group and pointwise convolutions. Later, Chen \etal proposed H-\ac{nerv}~\cite{chen2023hnerv} a hybrid neural representation for videos, advancing beyond standard \acp{inr} like \ac{nerv} and E-\ac{nerv} by incorporating learnable, content-adaptive embeddings. More recently, Kwan \etal introduced Hi\ac{nerv}~\cite{kwan2023hinerv}, a novel approach to learning-based video compression by leveraging hierarchical encoding and a sophisticated architecture combining bilinear interpolation, depth-wise convolutional, and \ac{mlp} layers. Finally, Maya \etal~\cite{maiya2023nirvana} employed a patch-wise autoregressive design through an \ac{inr} framework namely NIRVANA, enabling efficient scaling with video resolution and length, and introduces variable bitrate compression that adapts to inter-frame motion. 

\Ac{nerv} and related works~\cite{li2022nerv,bai2023ps,lee2022ffnerv,chen2023hnerv,kwan2023hinerv} require the training of networks with different architectures to achieve various \ac{rd} tradeoffs. They apply post-training quantization or \ac{qat}, and weight pruning to achieve compression. Although these approaches have demonstrated fast decoding capability, they still feature a relatively coarse decoder design, imposing limitations on the extraction of relevant video frame features and, consequently, restricting their rate-distortion performance.


One of the main challenges of implicit neural video representation lies in the ability to discern crucial information necessary for video representation, given the constraints imposed by the number of parameters in the decoder blocks. Another significant challenge is that current \ac{inr}-based models are not explicitly designed with architectural efficiency as a priority and lack the expressiveness needed to represent high-resolution video data at low distortion. This deficiency stems from ineffective decoding blocks, posing a challenge for video compression tasks. While existing approaches strive to enhance training and compression methods, there is a continued need to improve the representation capacity of the decoder network. Lastly, encouraging further exploration of \acp{inr} as an alternative to autoencoder-based \ac{nvc} is essential to overcome the limitations associated with conventional neural video compression methods. To address these challenges, we present three contributions outlined as follows:

\begin{itemize}
    \item We have developed a novel compact convolutional architecture for neural video representation, surpassing the representation capacity of state-of-the-art non-hybrid \ac{inr}-based video codecs. Fig.~\ref{HL_diag} illustrates a high-level diagram to provide a more comprehensive overview of the proposed framework.
    \item We have designed the \ac{nerv}++ decoder block as a simpler yet effective enhancement over the original \ac{nerv} decoder architecture. Our development is characterized by a \ac{scrb} that encompasses the traditional \ac{nerv} \ac{ub}. Furthermore, we have incorporated a bilinear interpolation layer to refine the feature representation capabilities of the decoder block. This design represents a technical leap forward, offering a more efficient solution for advanced video compression.
    \item We have conducted extensive experiments on key benchmark datasets for the video compression task. \ac{nerv}++ exhibits competitive qualitative and quantitative results compared to previous works and yields high-fidelity time-continuous reconstructions.
\end{itemize}
%
Thorough experimentation reveals the effectiveness of the \ac{nerv}++ decoder architecture, incorporating separable convolutions, residual connections, and bilinear interpolation layers. These experiments affirm that the \ac{nerv}++ framework delivers significant advances in compression efficiency.

\begin{figure*}[htb]
\centering
\includegraphics[width=1\textwidth]{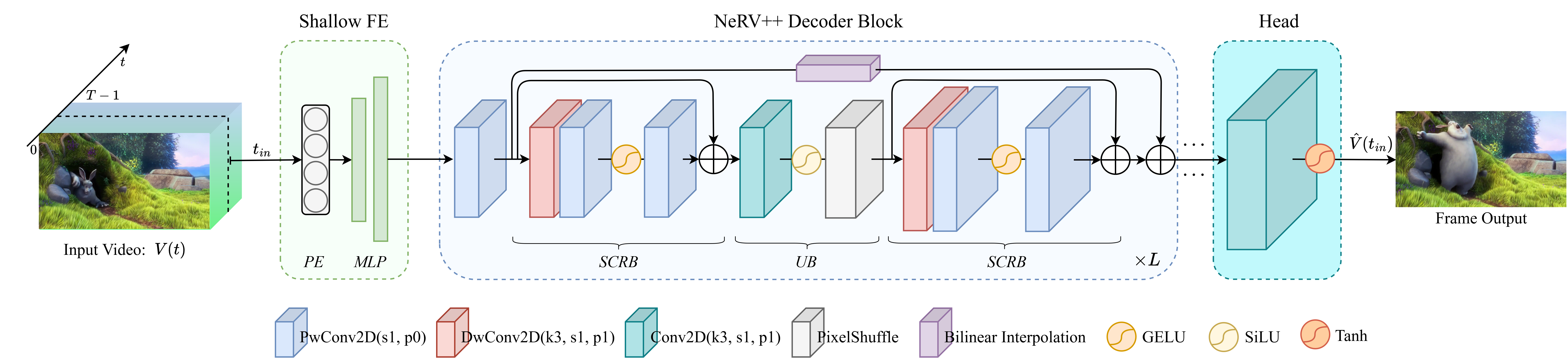}
\caption{Overall \acs{nerv}++ framework. We illustrate the video compression diagram of our \acs{nerv}++. PE for positional encoding, \ac{mlp} for multilayer perceptron, \acs{srcb} stands for the separable conv2d residual block, and UB for upsampling block.}
\label{ov_framework}
\end{figure*}

The rest of this paper is organized as follows. First, the proposed \ac{nerv}++ framework is described in detail in Section~\ref{ov_framework}. Next, we dedicate Section~\ref{result} to describing and analyzing the experimental results. Finally, Section~\ref{conclusion} concludes the paper and outlines some limitations of this work.

%
\section{Proposed NeRV++ Framework}
\label{method}
Our work comprises two proposals for improving neural representations for video compression: a more efficient neural network architecture for video compression (Sec.~\ref{nerv_arch}) and the formalization of the task as a \ac{rd} problem (Sec.~\ref{rd_perf}). Fig.~\ref{HL_diag} illustrates an overview of the proposed \ac{inr} for video compression, named \ac{nerv}++. Given a time coordinate $t_{in}$, the \ac{pe} with the \ac{mlp} decode the time coordinate and output a corresponding feature vector. This embedding is then forwarded to a set of sequential \ac{nerv}++ blocks to decode the frame at the corresponding time $t_{in}$. This overfitted model on the input video sequence is pruned, then quantized, and finally entropy-coded to produce the final compressed video bitstream.
%
%
\subsection{NeRV++ Overall Architecture}
\label{nerv_arch}
%
%
Given an input raw video ${V(t_{in}), t_{in} \in {0 \dots T} }$, our goal is to find a continuous representation for the video. The representation interprets an arbitrary time coordinate $t_{in}$ into an $RGB$ video frame. To achieve this, we introduce an enhanced implicit neural video representation, \ac{nerv}++. It enables time-continuous neural video representation and is parameterized by \acp{mlp} and \ac{nerv}++ decoder blocks, taking the form: $\hat{V}(t_{in})=f(t_{in})$, where $f$ is the proposed video representation defined by the encoded feature and network parameters. $t_{in}$ is the temporal coordinate, and $\hat{V}(t_{in})$ is the predicted $RGB$ frame at the instant $t_{in}$, as depicted in Fig.~\ref{HL_diag}.

The overall pipeline of the proposed solution is illustrated in Fig.~\ref{ov_framework}. The framework comprises three modular parts. First, the shallow feature extraction consists of a positional encoding operation and a small \ac{mlp}. Second, the \ac{nerv}++ decoder consists of a combination of separable conv2d residual blocks that sandwich the upsampling layers. We insert a bilinear interpolation skip connection between these layers for feature refinement. Finally, the convolutional head, coupled with a Tanh activation function, performs the final projection. This architecture enables the model to learn weights that achieve high-accuracy video representations and maintain low entropy, optimizing for compression efficiency.

Globally, the \ac{scrb} incorporates a series of architectural choices from a Swin Transformer~\cite{liu2021swin} while maintaining the network's simplicity as a standard \ac{convnet} without introducing any attention-based modules. These design decisions can be summarized as follows: macro design, ResNeXt's grouped convolution, inverted bottleneck, large kernel size, and various layer-wise micro designs. In Fig.~\ref{ov_framework}, we illustrate the \ac{scrb} block, where DwConv2D(.) refers to depthwise 2D convolution, PwConv2D(.) for point-wise 2D convolution, and \ac{gelu} for the activation function. Finally, it is essential to note that we propose to add a residual connection to simplify the learning process, enabling the training of much deeper networks by allowing gradients to flow through the architecture more effectively, thus improving performance without adding complexity.

\subsection{Model Compression Pipeline}
Model compression is typically achieved through pruning and quantization of network weights. Pruned models contain a majority of zeros and can be stored in sparse matrix formats for model size reduction. Alternatively, quantization works to reduce the number of bits needed to store each model weight, resulting in reduced disk space. As implicit neural networks represent the data using their model weights, data compression translates to model compression. We perform post-hoc pruning and quantization of the model because it has been found that \ac{qat} considerably increases training time for a relatively small improvement compared to \ac{ptq}.
\\\\
{\bf Weight pruning.}
In \ac{inr}-based methods~\cite{chen2021nerv,li2022nerv,bai2023ps,chen2023hnerv,lee2022ffnerv,kwan2023hinerv,maiya2023nirvana}, weight pruning is often employed to enhance model compression by globally zeroing weights with the smallest magnitudes, leveraging the principle that these contribute minimally to the model's output. Specifically, $L1$ unstructured pruning, a technique supported by the PyTorch library~\cite{prunpytorch}, applies this concept by targeting individual weights based on their $L1$ norm, promoting a sparse representation without adhering to the structure of channels or layers. Following the pruning process, fine-tuning is essential to recover or boost the model's performance. This stage adjusts the remaining weights to compensate for the loss induced by pruning, using a reduced learning rate and potentially fewer training epochs. This combined approach of pruning and fine-tuning facilitates the deployment of efficient, compact \ac{inr} models without significantly sacrificing accuracy, making it ideal for applications in resource-limited environments. We also evaluated our model with $L1$ unstructured pruning for fair comparisons, eliminating 20\% of the total convolutional weights.
%
\\\\
{\bf Weight quantization and entropy coding.}
In the domain of \ac{inr}-based video compression, weight quantization and entropy coding emerge as pivotal techniques for optimizing storage and transmission efficiency~\cite{chen2021nerv}. Weights quantization reduces the precision of the neural network's weights to a limited set of values, thereby decreasing the model's size and the computational complexity involved in processing video data. This step is crucial for fitting the neural representation within bandwidth constraints while maintaining acceptable video quality. Following quantization, entropy coding is applied to further compress the model by exploiting the statistical dependencies of the quantized weights, encoding them in a more compact form based on their frequency of occurrence. Techniques such as Huffman coding~\cite{moffat2019huffman} or arithmetic coding~\cite{rissanen1979arithmetic} are commonly used, efficiently mapping more frequent patterns to shorter codes. Together, weight quantization and entropy coding significantly enhance the compression ratio of \ac{inr}-based video compression methods, enabling high-quality video to be stored and transmitted with minimal resource usage. To achieve these outcomes, we utilized 8-bit post-training weight quantization coupled with Huffman entropy coding, further refining our compression efficiency.
%

\section{Results and Analysis}
\label{result}
\begin{table}[t]
\centering
\caption{PSNR (dB)$\uparrow$ performance on \ac{uvg} (1080p) dataset of non-hybrid \acp{inr} for video compression.}\label{psnr}
\adjustbox{max width=0.48\textwidth}{
\begin{tabular}{@{}l|l|cccccccc@{}}
\toprule
Size  & Video Codec  & Beauty & Bosph. & Honey. & Jockey & Ready. & Shake. & Yacht. & {\bf Average }\\
\midrule
\multirow{5}{*}{\rotatebox{90}{Small}}
& \acs{nerv}           &32.83&         32.20&         38.15&         30.30&         23.62&         33.24&         26.43&         30.97 \\
& E-\acs{nerv}         &33.13&         33.38&         \textbf{38.87}&         30.61&         24.53&         \textbf{34.26}&         26.87&         31.66 \\
& PS-\acs{nerv}        &32.94&         32.32&         38.39&         30.38&         23.61&         33.26&         26.33&         31.03 \\
& \acs{nerv}++         &33.48&         34.03&         38.65&         31.87&         24.91&         33.97&         27.37&         32.04 \\
& \acs{nerv}*++        &\textbf{33.57}&         \textbf{34.59}&         38.76&         \textbf{32.46}&         \textbf{25.29}&         34.11&         \textbf{27.65}&         \textbf{32.35} \\
\bottomrule
\multirow{5}{*}{\rotatebox{90}{Medium}}
& \acs{nerv}           &33.67&         34.83&         39.00&         33.34&         26.03&         34.39&         28.23&         32.78 \\
& E-\acs{nerv}         &\textbf{33.97}&         35.83&         \textbf{39.75}&         33.56&         26.94&         \textbf{35.57}&         28.79&         33.49 \\
& PS-\acs{nerv}        &33.77&         34.84&         39.02&         33.34&         26.09&         35.01&         28.43&         32.93 \\
& \acs{nerv}++         &33.90&         35.71&         39.08&         34.33&         26.79&         34.52&         28.79&         33.30 \\
& \acs{nerv}*++        &\textbf{33.97}&         \textbf{36.16}&         39.15&         \textbf{34.74}&         \textbf{27.20}&         34.60&         \textbf{29.06}&         \textbf{33.55} \\
\bottomrule
\multirow{5}{*}{\rotatebox{90}{Large}}
& \acs{nerv}           &34.15&         36.96&         39.55&         35.80&         28.68&         35.90&         30.39&         34.49 \\
& E-\acs{nerv}         &34.25&         37.61&         \textbf{39.74}&         35.45&         29.17&         \textbf{36.97}&         30.76&         34.85 \\
& PS-\acs{nerv}        &\textbf{34.50}&         37.28&         39.58&         35.34&         28.56&         36.51&         30.28&         34.58 \\
& \acs{nerv}++         &34.25&         37.47&         39.46&         36.22&         29.23&         35.72&         30.78&         34.72 \\
& \acs{nerv}*++        &34.28&         \textbf{37.77}&         39.48&         \textbf{36.45}&         \textbf{29.59}&         35.83&         \textbf{30.99}&         \textbf{34.91} \\
\bottomrule
\end{tabular}%
}
\end{table}
\subsection{Experimental Setup}
\label{setup}
{\bf Baselines.}
We compare our solution with the benchmark \ac{inr}-based video codecs, including \acs{nerv}~\cite{chen2021nerv}, E-\acs{nerv}~\cite{li2022nerv}, and PS-\acs{nerv}~\cite{bai2023ps}.
\\\\
{\bf Datasets.}
We assessed the performance of our method using seven 1080p videos from the widely utilized \ac{uvg} video dataset~\cite{mercat2020uvg}, four 720p videos from the MCL\_JVC dataset~\cite{wang2016mcl}, and one 720p Bunny video from the scikit-video test dataset~\cite{bunny}. This evaluation aimed to compare our approach with other neural field-based video compression methods. The chosen datasets encompass a diverse range of videos, spanning from nearly static scenes to fast-moving sequences.
\\\\
{\bf Implementation details.}
We implemented all models on PyTorch, and the experimental study was carried out on an RTX 5000 Ti GPU and an Intel(R) Xeon(R) W-2145 @ 3.70GHz CPU. All models were overfitted on the same video sequences from the considered datasets with 300 epochs using the ADAM optimizer with parameters $\beta_1=0.9$ and $\beta_2=0.999$. The initial learning rate is set to $5 \times 10^{-4}$ with a cosine learning rate scheduler. We used a weighting between \ac{mae} and (1-\acs{ssim}) to formulate the loss function in $RGB$ color space. This loss function is expressed as follows, where $\mathbf{x}$ and $\mathbf{\hat{x}}$ stand for predicted and target frames: $$\mathcal{L} = 0.7 \times MAE(\mathbf{{x}}, \mathbf{\hat{x}}) + 0.3 \times [1 - SSIM(\mathbf{{x}}, \mathbf{\hat{x}})]$$ To cover a wide range of rate and distortion points, for our proposed method and baseline, we trained four models (xsmall, small, medium, and large configurations) per video for all considered neural codecs.
\\\\
{\bf Evaluation.}
The performance metrics are \ac{psnr} and \ac{ms-ssim} at several coding bitrates. We also calculate the BD-rate savings \cite{bjontegaard2001calculation}, and compare the model size and inference time to measure the model's efficiency and complexity.

\subsection{Rate-Distortion Coding Performance}
\label{rd_perf}
In this study, we present an evolution of the \ac{nerv}++ architecture, denoted as \ac{nerv}*++, which incorporates a modification to its primary \ac{scrb} by deepening the feature representation in the second layer. This adjustment is aimed at enhancing the model's ability to capture intricate details.

\begin{figure}[htb]
\centering
\includegraphics[width=0.5\textwidth]{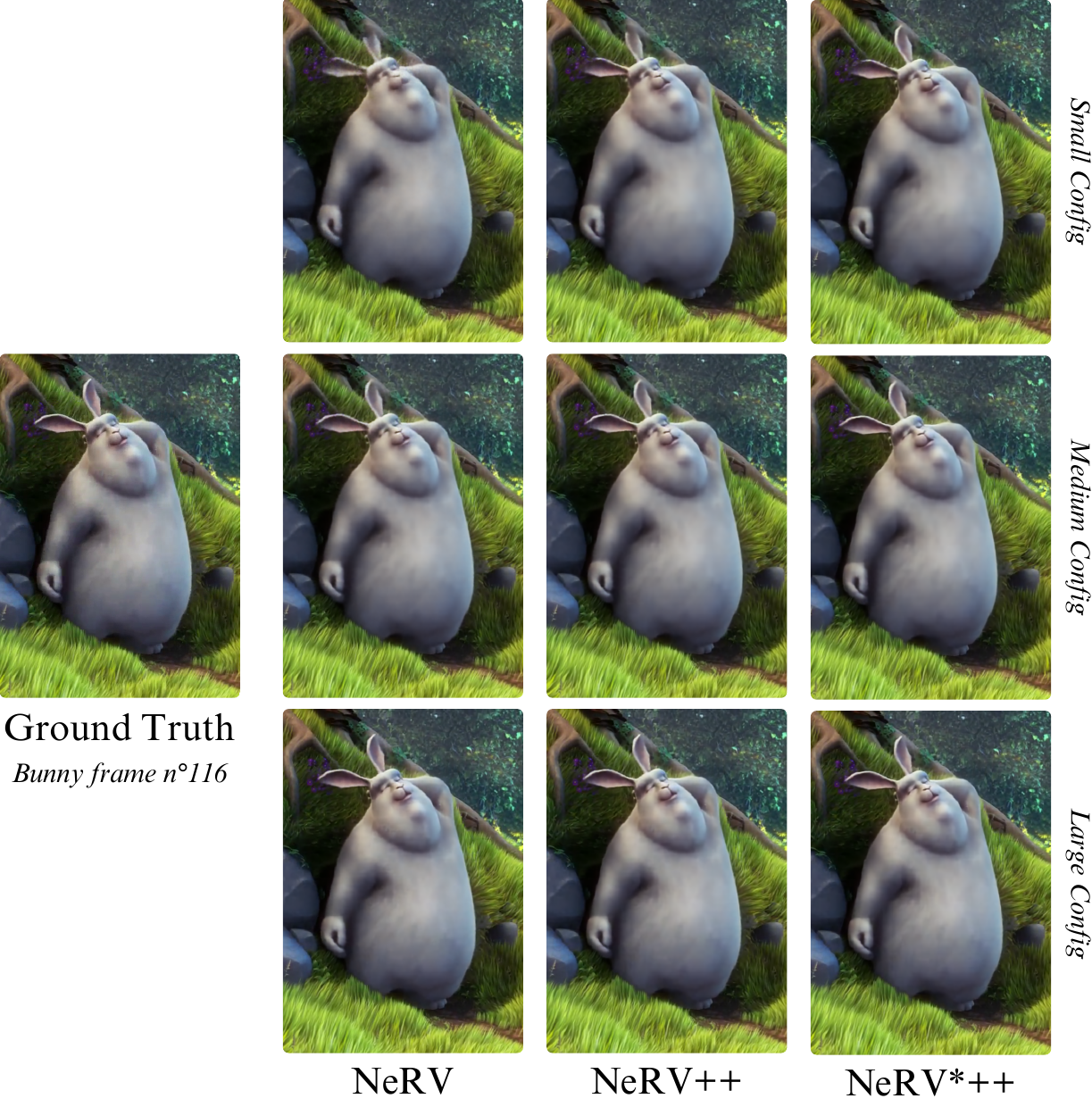}
\caption{Visualization of the reconstructed frame number 116 from the Bunny dataset.}
\label{qual_analysis}
\end{figure}

To showcase the compression efficiency of our proposed approach, \ac{nerv}++, and its evolution, \ac{nerv}*++, we give the \ac{psnr} results in Table~\ref{psnr} across the \ac{uvg} (1080p) dataset, comparing with state-of-the-art non-hybrid \ac{inr}-based video codecs. On average, \ac{nerv}*++ achieves a $0.86dB$ enhancement in \acs{psnr} compared to \ac{nerv} across all \ac{uvg} videos. In all cases, our method produces visually sharper videos than \ac{nerv} and PS-\ac{nerv} for equal or smaller bitrates, as represented in Fig.~\ref{qual_analysis}. Videos such as "ShakeNDry" and "Honeybee" often present challenges to compression algorithms relying on motion estimation, such as \ac{hevc} or \ac{ssf}. On the other hand, for "YachtRide," motion-based descriptors compress the scene well, as most motion vectors are parallel. While our method clearly outperforms \ac{nerv} and PS-\ac{nerv} in this scene, E-\ac{nerv} shows close results for medium and large model configurations.

Considering Table~\ref{bdrate_psnr_msssim}, \ac{nerv}++ and \ac{nerv}*++ significantly outperform the \ac{nerv} approach on both evaluated metrics, demonstrating the advantages of the \ac{nerv}++ decoder block architecture and the well-engineered design of the \ac{scrb}. This establishes new non-hybrid state-of-the-art results for implicit neural video representation.

\begin{table}[t]
\centering
\caption{BD-rate$\downarrow$ performance of \acs{nerv}++, and \acs{nerv}*++ compared to \acs{nerv}. Vid.$\{1 \dots 4\}$ are four 720p videos taken from MCL\_JVC dataset. Bunny is a 720p video from scikit-video test dataset.}\label{bdrate_psnr_msssim}
\adjustbox{max width=0.48\textwidth}{
\begin{tabular}{@{}l|cccccc@{}}
\toprule
Video Codec     & Bunny & Vid.1 & Vid.2 & Vid.3 & Vid.4 & {\bf Average }\\
\midrule
  & \multicolumn{5}{c}{BD-rate (\acs{psnr} dB)$\downarrow$}           \\
\midrule
\acs{nerv}++   & \textbf{-27.98\%}  & -21.46\% & -27.24\% & \textbf{-40.62\%} & \textbf{-17.50\%} & \textbf{-26.96\%} \\
\acs{nerv}*++  & -25.68\%  & \textbf{-24.69\%} & \textbf{-30.55\%} & -26.47\% & -16.59\% & -24.796\% \\
\midrule & \multicolumn{5}{c}{BD-rate (\acs{ms-ssim} dB)$\downarrow$}  \\
\midrule
\acs{nerv}++   & \textbf{-29.99\%}  & -20.37\% & -24.73\% & \textbf{-33.35\%} & \textbf{-15.26\%} & \textbf{-24.74\%}  \\
\acs{nerv}*++  & -25.26\%  & \textbf{-22.05\%} & \textbf{-28.18\%} & -26.66\% & -13.69\% & -23.17\%  \\
\bottomrule
\end{tabular}%
}
\end{table}

\subsection{Models Scaling Study}
\label{modelscaling}
In \ac{inr}-based approaches, the decoding strategy, which relies on executing a forward pass for each frame independently, significantly enhances the potential for parallel processing in the decoding process, thereby improving efficiency.
We evaluated the decoding complexity of the proposed method and the \ac{nerv} baseline by averaging decoding time across 132 frames at 720p resolution, encoded on average at 0.005 bpp. Table~\ref{complexity} presents the video codec complexity features, displaying the decoding latency (fps), \acp{mac} per pixel, and parameter counts for our method compared to the baseline \ac{nerv}. Finally, we note that the models were executed using PyTorch on a workstation with one RTX 5000 Ti \ac{gpu}.
According to Tables~\ref{bdrate_psnr_msssim} and \ref{complexity}, \ac{nerv}++ can achieve significantly better \ac{rd} performance with a lower \acp{mac} per pixel and number of parameters but higher latency on the \ac{gpu} compared to \ac{nerv}.
Our findings suggest the potential for significantly reducing decoding times by optimizing the batch size, given adequate \ac{gpu} memory. This aligns with the observations of Chen \etal~\cite{chen2021nerv} that \ac{inr}-based video compression methods outperform alternative models in decoding efficiency.
\begin{table}[t]
\centering
\caption{Average decoding latency across 132 frames at 720p resolution, encoded on average at 0.05 bpp.
}\label{complexity}
\adjustbox{max width=0.48\textwidth}{%
\begin{tabular}{@{}l|ccc@{}}
\toprule
Video Codec & Latency(fps)$\uparrow$ & \#\acp{mac} ppx$\downarrow$ & \#parameters(M)$\downarrow$ \\
\midrule
\ac{nerv}           & 24.89 & 277.8 & 5.96  \\
\ac{nerv}++ (ours)  & 15.53 & 256.3 & 5.83  \\
\ac{nerv}*++ (ours) & 10.54 & 366.8 & 6.07  \\
\bottomrule
\end{tabular}%
}
\end{table}

\section{Conclusion and Limitations}
\label{conclusion}
%
In this study, we introduced the \ac{nerv}++ model, an advanced refinement of the \ac{nerv} framework aimed at enhancing video compression efficiency. Our innovation lies in the development of a \ac{nerv}++ decoder block, which integrates a \ac{scrb} around the conventional \ac{nerv} upsampling unit, supplemented by a bilinear interpolation layer for improved feature refinement. This configuration not only enhances the architecture but also significantly elevates the performance on 12 videos from \ac{uvg}, MCL\_JVC, and Bunny datasets, positioning \ac{nerv}++ as a potent solution in the field of video compression technology.
While the current findings are encouraging, the practical implementation of neural representations in video compression as efficient encoders necessitates further investigation into more economical entropy-modeling techniques to enhance encoding efficiency. Additionally, the model complexity and decoding latency still need improvement. It is also possible for the quantized model to outperform the full-precision 32-bit one if the quantized model is initialized with a bigger and deeper model. Besides quantization, other compression techniques like knowledge distillation may also enhance the model compression pipeline.
%

\small
\bibliographystyle{IEEEbib}
\bibliography{egbib}

\end{document}